\documentclass[10pt]{article}
\usepackage[a4paper, total={6.5in, 9in}]{geometry}
\usepackage{hyphenat}
\usepackage[cmex10]{amsmath} 
\usepackage{amssymb}         
\usepackage{amsfonts}        
\usepackage{bm}  
\usepackage{bbm}
\usepackage{graphicx}
\usepackage{longtable}
\usepackage{subcaption} 
\usepackage{float} 
\usepackage{microtype}
\usepackage{optidef}
\usepackage{comment}
\usepackage[numbers]{natbib}
\usepackage{color}
\usepackage[ruled,lined,noend,linesnumbered]{algorithm2e}
\SetKwProg{Init}{Initialization}{}{}
\usepackage{hyperref}

\allowdisplaybreaks

\begin{document}

\title{Energy Efficient Task Offloading in UAV-Enabled MEC Using a Fully Decentralized 
Deep Reinforcement Learning Approach}

\author{%
  Hamidreza Asadian-Rad\thanks{H. Asadian-Rad, H. Soleimani, and S. Farahmand are with the Department of Electrical Engineering, Iran University of Science and Technology (IUST), Tehran 1684613114, Iran 
  (e-mail: 77asadian@gmail.com; {hsoleimani@iust.ac.ir}; shahrokhf@iust.ac.ir).}, \and
  Hossein Soleimani$^\ast$\thanks{Corresponding author: Hossein Soleimani.}, \and
  Shahrokh Farahmand$^\ast$
}

\maketitle

\begin{abstract}
Unmanned aerial vehicles (UAVs) have been recently utilized in multi-access edge computing (MEC) as edge servers. Given their agility and non-terrestrial nature, UAVs provide strong line of sight (LoS) channels to mobile users, and can service areas with non-existent or malfunctioning  infrastructure. It is desirable to design UAVs' trajectories and user to UAV assignments to ensure satisfactory service to the users and energy efficient operation simultaneously. The posed optimization problem is challenging to solve because: (i) The formulated problem is non-convex, (ii) Due to the mobility of ground users, their future positions and channel gains are not known in advance, (iii) Local UAVs' observations should be communicated to a central entity that solves the optimization problem. The (semi-) centralized processing leads to communication overhead, communication/processing bottlenecks, lack of flexibility and scalability, and loss of robustness to system failures. To simultaneously address all these limitations, we advocate a fully decentralized setup with no centralized entity. Each UAV obtains its local observation and then communicates with its immediate neighbors only. After sharing information with neighbors, each UAV determines its next position via a locally run deep reinforcement learning (DRL) algorithm. None of the UAVs need to know the global communication graph. Two main components of our proposed solution are (i) Graph attention layers (GAT), and (ii) Experience and parameter sharing proximal policy optimization (EPS-PPO). Our proposed approach eliminates all the limitations of semi-centralized MADRL methods such as MAPPO and MA deep deterministic policy gradient (MADDPG), while guaranteeing a better performance than independent local DRLs such as in IPPO. Numerical results reveal notable performance gains in several different criteria compared to the existing MADDPG algorithm, demonstrating the potential for offering a better performance, while utilizing local communications only.
\end{abstract}


\section{Introduction}

The proliferation of new applications that require demanding processing power, to deliver high-quality services with low latency, has imposed a great burden on the computing resources of current user devices. As examples of such applications, one can point to autonomous driving, smart cities, and online gaming \cite{zhou2020mobile}. Multi-access edge computing (MEC), has emerged as a key solution, which helps users to process their data with minimal latency by offloading it to edge servers \cite{mao2017survey}. On the other hand, unmanned aerial vehicles (UAVs) are highly suited to applications that demand dynamic and adaptive deployment of processing servers. Due to their operational flexibility, high mobility, and high probability of line-of-sight (LoS) links to the users, UAVs are very desirable as MEC servers. They offer the potential to improve connectivity to ground users, by simultaneously optimizing their trajectories and energy consumption \cite{mozaffari2019tutorial}, \cite{sekander2018multi}, \cite{mozaffari2017mobile}, \cite{wu2018joint}, \cite{xiao2016enabling} to ensure energy-efficient seamless operation. These characteristics render UAVs even more valuable and reliable in scenarios where fixed ground edge servers are not available or are damaged \cite{zhao2021predictive}, \cite{pham2020survey}.

\begin{table}[t!]
\centering
\caption{Comparison Between Our Work and the Existing Literature}
\label{tab:comparison}
\resizebox{\textwidth}{!}{%
\begin{tabular}{|c|c|c|c|c|c|c|c|c|}
\hline
\textbf{Reference} & \multicolumn{3}{|c|}{\textbf{Algorithm Category}} & \multicolumn{3}{|c|}{\textbf{Learning Type}} & \multicolumn{2}{|c|}{\textbf{Considered Features}} \\
\hline
                   & Centralized & Semi-Distributed & Decentralized & DRL & Non-DRL & Multi-Agent & Energy & Task Offloading \\
\hline
\cite{hu2018joint, qian2019user} & \checkmark &   &   &   & \checkmark &   &   & \checkmark \\
\hline
\cite{wang2021task, liu2022resource, li2020energy, yang2022online, ji2020energy, xu2021uav, dusit} & \checkmark &   &   &   & \checkmark &   & \checkmark & \checkmark \\
\hline
\cite{song2022evolutionary} & \checkmark &   &   & \checkmark &   &   & \checkmark & \checkmark \\
\hline
\cite{lin2023pddqnlp,30} & \checkmark &   &   & \checkmark & \checkmark &   & \checkmark & \checkmark \\
\hline
\cite{wang2021deep, wang2020multi, chen2022deep} &   & \checkmark &   & \checkmark & \checkmark &   & \checkmark & \checkmark \\
\hline
\cite{liu2024deep, wang2024decentralized,H4Rob, H7} &   & \checkmark &   & \checkmark &   & \checkmark & \checkmark & \checkmark \\
\hline
\cite{li2023computing, he2023fairness} &   & \checkmark &   & \checkmark & \checkmark & \checkmark & \checkmark & \checkmark \\
\hline
\cite{29},~Our Algorithm &   &   & \checkmark & \checkmark &   & \checkmark & \checkmark & \checkmark \\
\hline
\end{tabular}%
}
\end{table}

Existing literature on UAV-assisted MEC problems, can be divided into two main categories: Single UAV problems \cite{hu2018joint, qian2019user, wang2021task, liu2022resource, li2020energy, yang2022online, ji2020energy, xu2021uav, song2022evolutionary, lin2023pddqnlp, liu2024deep}, and Multiple UAV problems \cite{30,wang2021deep, wang2020multi, chen2022deep, wang2024decentralized,H4Rob,H7, li2023computing, he2023fairness} . In single UAV problems, a UAV is either assisting ground edge servers \cite{xu2021uav}, \cite{song2022evolutionary}, or it is the stand alone edge server \cite{hu2018joint, qian2019user, wang2021task, liu2022resource, li2020energy, yang2022online, ji2020energy, lin2023pddqnlp, liu2024deep}. In all single UAV setups, UAV trajectory is a main design parameter. In addition, users to UAV associations, and optimal allocation of computation and communication resources are also design parameters. Optimization objectives include maximizing total number of offloaded tasks or bits \cite{qian2019user}, maximizing energy-efficiency \cite{li2020energy,xu2021uav,lin2023pddqnlp}, minimizing energy-consumption \cite{wang2021task, liu2022resource, yang2022online, ji2020energy,liu2024deep}, and minimizing task processing delays \cite{hu2018joint}. A multi-objective optimization is investigated by \cite{song2022evolutionary}, where three of the aforementioned objectives are simultaneously optimized. A major assumption in most single UAV works is that users are either fixed during the whole episode that UAV trajectory is being designed \cite{hu2018joint, qian2019user, wang2021task, liu2022resource,ji2020energy, xu2021uav, song2022evolutionary,lin2023pddqnlp, liu2024deep} or their movement is completely known by the optimization algorithm \cite{li2020energy}. This simple assumption may not necessarily hold in practice and becomes a main limitation of the aforementioned works. For a single UAV, only \cite{yang2022online} treats the users' movement as random and not known in advance. This work utilizes Lyapunov optimization which is a main stochastic optimization technique for dynamic environments. Most aforementioned works utilize classic optimization techniques, in spite of the challenging nature of the formulated optimization problems. However, few works  \cite{song2022evolutionary,lin2023pddqnlp,liu2024deep} utilize centralized deep reinforcement learning (DRL) to solve the single UAV problem. 

The preferred solution changes considerably for multiple UAV setups, where different instances of DRL and multi-agent DRL (MADRL) are utilized in most references \cite{30, wang2021deep, wang2020multi, chen2022deep,wang2024decentralized, H4Rob, H7, he2023fairness, 29}. There exists a reference that solves the multiple UAV MEC problem via classic optimization \cite{dusit}, but it assumes fixed users at known positions and a centralized optimizing entity. It should be noted that there are many UAV-enabled MEC setups, where UAV trajectory is assumed fixed and given. Then, optimization is carried out over users to UAVs associations and computation and communication resource allocation. We do not cite these references, as UAV trajectory design is our main target. However, they can be found in \cite{dusit} for the interested reader. In multiple UAV problems, the objective is either to minimize total energy consumption \cite{30, wang2021deep, chen2022deep,H4Rob,he2023fairness,29}, maximize energy efficiency \cite{wang2020multi,wang2024decentralized,li2023computing}, minimize latency \cite{30,chen2022deep}, or maximize the number of offloaded tasks \cite{H7}. These approaches rely either completely \cite{30, wang2021deep, chen2022deep, li2023computing}, or partially \cite{wang2020multi,wang2024decentralized,H4Rob,H7,he2023fairness} on a centralized entity to optimize the system parameters. In the centralized approaches, a central entity with the full knowledge of environment state, runs the DRL algorithm and communicates the optimized decisions to the UAVs. Semi-centralized approaches mostly rely on the centralized learning decentralized execution (CLDE) framework and employ algorithms such as multi-agent deep deterministic policy gradient (MADDPG) or multi-agent proximal policy optimization (MAPPO). The existence of a centralized entity in both of the aforementioned approaches brings about several limitations. These UAV networks lack scalability and flexibility, suffer from computation and communication bottlenecks, and are very sensitive to central entity failure. Indeed, these networks suffer from a single point of failure issue, meaning that if the centralized entity breaks down the whole network and its performance are heavily affected. The only fully decentralized work in the literature is \cite{29}, which utilizes the independent PPO (IPPO) algorithm proposed in \cite{H2}. The limitations of IPPO will be alluded to later in the introduction. Table I provides a summary of the available references and their proposed solutions.

In this work, we advocate a fully decentralized setup where no central entity exists in the network. Each UAV only observes those users that are in its coverage range, and it can communicate with only those UAVs that are in its communication range. Given this fully decentralized nature, none of the UAVs have a global view of the environment state. Instead, they only collect local observations from nearby users and UAVs. Still, the goal is to solve a certain global optimization problem in a decentralized fashion. Our proposed setup is both scalable and flexible, does not suffer from computation and communication bottlenecks, and is robust to system failures. This means that if any UAV breaks down or returns to base for recharging, the network can still operate satisfactorily. The main challenge for this setup, is for each individual UAV to make good decisions with limited local observations only. The simplest such method is IPPO, where every UAV runs its local DRL algorithm without communicating with other agents \cite{H2}. It was demonstrated that while IPPO performs worse than MAPPO, the difference in performance is not staggering \cite{H2}. This performance gap arises mainly because IPPO does not allow communication between agents which is a necessity for performance improvement through coordination and cooperation. To address this issue, we advocate the incorporation of graph neural network (GNN) into the MEC framework. By using GNN, and specifically graph attention network (GAT) \cite{velivckovic2017graph}, we enable local communication between nearby UAVs. Through these local information sharing, the global network information is diffused in the network, leading to a more accurate individual view of the environment global state by each UAV. In addition to GAT layers, we utilize an experience and parameter sharing PPO (EPS-PPO) algorithm that shares both replay buffers and PPO actor-critic networks' weights with the immediate neighbors to improve performance. To the best of our knowledge, setting aside \cite{29}, no such fully decentralized method has been proposed for UAV-enabled MEC systems. 

Utilization of GNN, and GAT to be more specific, for MADRL is gaining popularity across many domains and applications. For example, in persistent monitoring of an area by multiple robots, a combination of GAT and PPO has been applied in \cite{H5,H6}. However, the communication graph between robots is assumed to be fully-connected \cite{H5} which is not usually the case, or MAPPO is utilized \cite{H6}, which requires a central entity. GNNs have been proposed for multiple UAVs serving as base stations also \cite{H3}. The UAVs communicate data to and from ground users, and the goal is to determine UAVs trajectories and users to UAV associations in such a way that a fairness-incorporated sum-rate is maximized. Unlike our proposed approach, \cite{H3} still requires the central entity to perform centralized training decentralized execution for a multi-agent deep Q network (DQN). Our proposed approach is inspired by the work in \cite{pu2022attention}. However, the approach in \cite{pu2022attention} is overly complicated and very complex to implement. Thus, we omitted certain blocks from their proposed architecture, to reach a computationally efficient design, while still reaping the benefits of their fully decentralized approach. Our posed optimization problem aims to minimize the total energy consumption for both UAVs and users, while maximizing the number of processed tasks. Our design parameters are UAV trajectories and user to UAVs associations. To solve the problem, we utilize the aforementioned fully decentralized method. The main contributions of this paper are summarized as follows:
\begin{itemize}
    \item
        Unlike existing UAV-enabled MEC literature, We assume a fully-decentralized setup, where UAVs have limited user coverage and communication range. They collect local observations, and are only allowed to communicate with immediate neighbors. No central entity exists. Our proposed setup entails robustness to node failures, elimination of communication and computation bottlenecks, and guarantees flexibility and scalability. 
    \item 
        Our proposed solution utilizes GAT to allow for communication between immediate neighbors. The GAT is operating in conjunction with EPS-PPO that shares experiences and actor-critic networks' weights between immediate neighbors. While, the algorithm runs locally on every UAV, communication between neighbors through GAT and EPS-PPO leads to dissemination of all UAVs information inside the network. Thus, UAVs individual decisions are made with a better awareness of the true global environment state. 
    \item
        Numerical evaluations reveal a double advantage compared to the semi-decentralized MADDPG benchmark. While our proposed approach does not suffer from the limitations of having the central entity in MADDPG, it performs better than MADDPG in certain criteria as well. 
\end{itemize}

The rest of this paper is organized as follows. 
In Section~\ref{sec:system_model} we introduce the system model and formulate the optimization problem. 
In Section~\ref{sec:algorithm} our fully decentralized MADRL approach is proposed. 
Section~\ref{sec:results} presents the numerical results, and conclusions are drawn in Section~\ref{sec:conclusions}.

\begin{figure}[t!]
    \centering
    \includegraphics[width=0.85\textwidth]{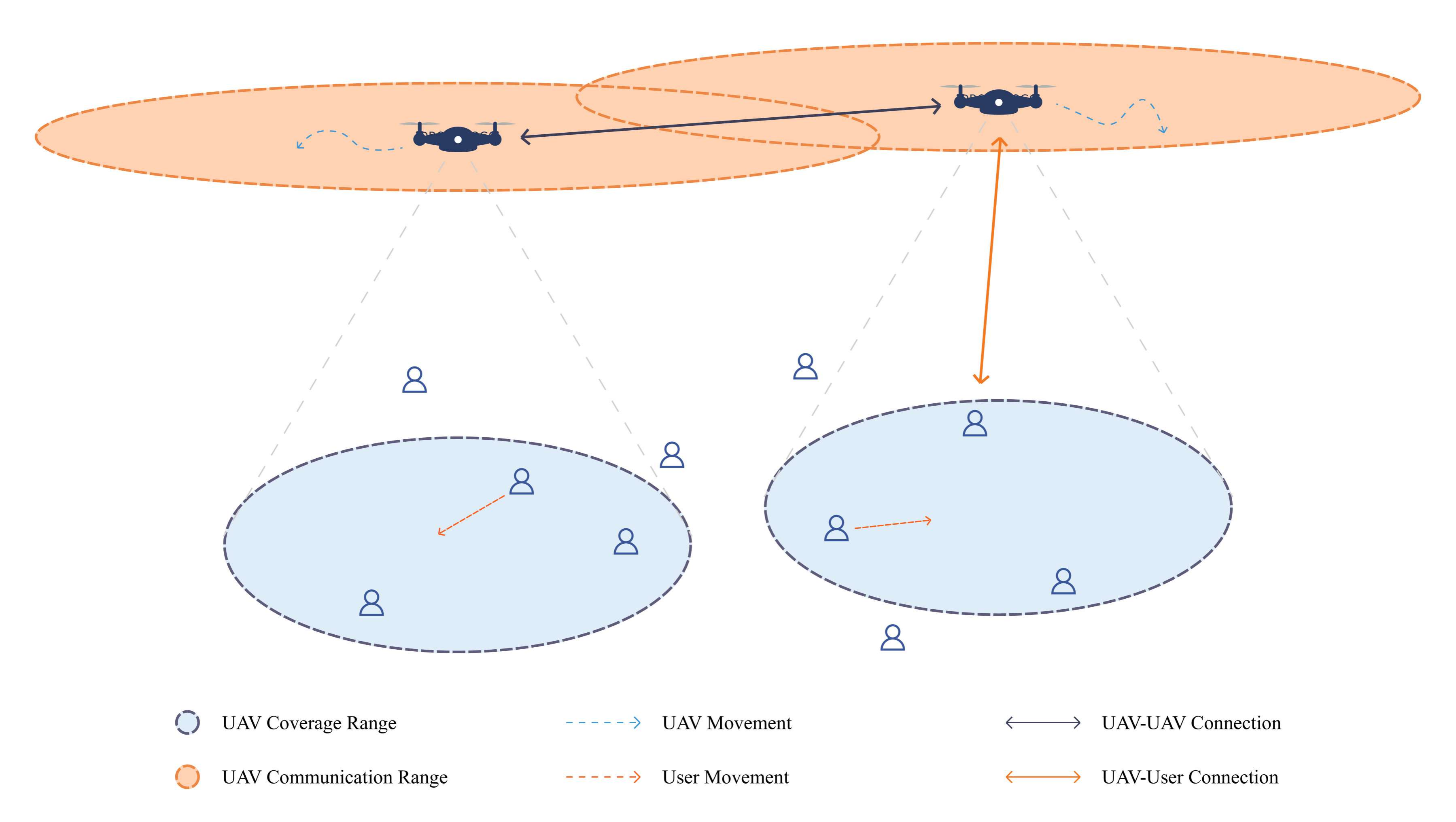}
    \caption{System Model of the Proposed Method.}
    \label{fig:system_model}
\end{figure}

\begin{longtable}{|p{2.8cm}|p{11.5cm}|} 
\caption{List of Main Notations} \label{tab:notations} \\

\hline
\textbf{Notation} & \textbf{Description} \\
\hline
\endfirsthead

\hline
\textbf{Notation} & \textbf{Description} \\
\hline
\endhead

$n$, $N$, $\mathcal{N}$ & Index, number, and set of users \\
\hline
$m$, $M$, $\mathcal{M}$ & Index, number, and set of UAVs \\
\hline
$t$, $T$, $\mathcal{T}$ & Index, number, and set of time slots \\
\hline
$\mathbf{q}_{m}(t)$ & UAV $m$'s coordinates at time slot $t$ \\
\hline
$\mathbf{q}_{n}(t)$ & User $n$'s coordinates at time slot $t$ \\
\hline
$\mathbf{v}_{m}$ & Velocity of UAV $m$ \\
\hline
$\mathbf{v}_{n}$ & Velocity of user $n$ \\
\hline
$d_{n,m}$ & Distance between user $n$ and UAV $m$ \\
\hline
$d_{m,m\prime}$ & Distance between UAV $m$ and UAV $m\prime$ \\
\hline
$R_{\rm cov}$ & UAV maximum coverage range \\
\hline
$R_{\rm com}$ & UAV maximum communication range \\
\hline
$\delta_{n, m}(t)$ & Coverage indicator for user $n$ at time slot $t$ \\
\hline
$\alpha_{n,m,t}$ & Offloading decision for user $n$ to UAV $m$ at time $t$ \\
\hline
$h_{n,m}(t)$ & Channel gain between user $n$ and UAV $m$ \\
\hline
$G_{0}$ & Power gain at 1 meter reference distance \\
\hline
$r_{n,m}(t)$ & Achievable data rate for user $n$ at time $t$ \\
\hline
$B$ & Uplink available bandwidth \\
\hline
$P_{n}$ & User $n$ transmit power \\
\hline
$\sigma_{m}$ & Noise power at UAV $m$ receiver \\
\hline
$S_{u,n}$ & Data size of task $u$ for user $n$ \\
\hline
$C_{u,n}$ & CPU cycles required for task $u$ of user $n$ \\
\hline
$\tau_{n,u,m}$ & Transmission time for task $u$ to UAV $m$ \\
\hline
$P_m^r$ & Power consumption of UAV $m$’s receiver circuits \\
\hline
$E_{n,u,m}^{t}(t)$ & Energy to send task $u$ of user $n$ to UAV $m$ \\
\hline
$E_{m}^{h}(t)$ & Hovering energy for UAV $m$ \\
\hline
$E_{m}^{f}(t)$ & Flying energy for UAV $m$ \\
\hline
$E_{n,u,m}^{p}(t)$ & Energy to process task $u$ of user $n$ at UAV $m$ \\
\hline
$\kappa_m$ & Energy per CPU cycle at UAV $m$ \\
\hline

\end{longtable}

\section{System Model and Problem Formulation}
\label{sec:system_model}
We assume a scenario where $M$ UAVs and $N$ users are randomly distributed within a 
square-shaped area of length $L$. The set of UAVs is denoted by $\mathcal{M}=\left\{1,2, 
\ldots, M\right\}$, where each UAV is equipped with a MEC server and flies at a constant 
altitude to serve nearby users. The set $\mathcal{N}=\left\{1,2, \ldots, N\right\}$ 
represents the users. Each computational task is processed within a time slot $t$, where 
$t\in \mathcal{T}=\left\{1,2, \ldots, T\right\}$. Each run through $\mathcal{T}$ is 
called an episode. During each episode, users move within the designated area. It is 
assumed that each user has a task buffer. After the first task is processed by one of the 
UAVs, the second task is read from the buffer and communicated to a UAV for processing. 
Every UAV can only serve those users that are located in its limited coverage 
area of radius $R_{\rm cov}$. Furthermore, every UAV can only communicate with its neighboring UAVs located in 
its limited communication range of radius $R_{\rm com}$. Subsequently, every UAV has a partial and limited view 
of available tasks and other UAVs observations and decisions, as it does not see the 
users and UAVs out of its coverage/communication range. System setup is depicted in Fig. 
\ref{fig:system_model}. System parameters are summarily presented in Table II. Furthermore, we assume that 
users do not have the capability to process their tasks locally. Hence, they completely 
rely on the swarm of UAVs to process their tasks. 

Each user's location is defined by $\mathbf{q}_n(t) = [x_n(t), y_n(t)]^T$, while the velocity of each user is given by $\mathbf{v}_n$ which belongs to $\mathbb{R}^2$ as well. Velocity may change at every time step. The position of user $n$ at time slot $t+1$ is given by
\begin{equation}
    \mathbf{q}_n(t+1) = \mathbf{q}_n(t) + \mathbf{v}_n(t) \Delta t,
\end{equation}
where $\Delta t$ is the duration of one time slot. We assume a simple bounce for the user if it hits one of the borders of the designated square area. Similarly, the position of a UAV is described by $\mathbf{q}_m(t) = [x_m(t), y_m(t), H]^T$. We assume that UAVs operate at a fixed altitude $H$ and can travel at a limited velocity within each time slot, as specified by $V_{max}$. The position of a UAV is updated dynamically according to
\begin{equation}
    \mathbf{q}_m(t+1) = \mathbf{q}_m(t) + \mathbf{v}_m(t) \Delta t,
\end{equation}
where $\mathbf{v}_m(t)$ is in $\mathbb{R}^3$ and its third element equals zero, as the altitude $H$ is constant. Similar to users, UAVs can not leave the designated area and they will remain still if they hit any of the borders. The distance between user $n$ and UAV $m$ at time slot $t$ is denoted by $d_{n, m}(t)$. This distance should be less than $R_{\rm cov}$, so that the user $n$ can be served by UAV $m$. To indicate the presence of a user within a UAV's coverage area, we define the indicator $\delta_{n,m}$ as follows:
\begin{equation}
    \delta_{n, m}(t)= \begin{cases}1, & \text { if } d_{n, m}(t) \leq R_{\rm cov} \\ 0, & \text { otherwise }\end{cases},
\end{equation}
where
\begin{equation}
    d_{n, m}(t) = \sqrt{(x_n(t) - x_m(t))^2 + (y_n(t) - y_m(t))^2 + H^2}
\end{equation}
We consider an episodic setup, where UAVs keep processing tasks until either all tasks are processed, or $T$ time slots have passed, whichever occurs sooner. As for tasks' details, we assume user $n$ has exactly $U_n$ tasks in its buffer at the beginning of an episode. No further tasks are generated during the episode. Each task $u$ is characterized by the tuple $[S_{n,u}, C_{n,u}]$, where $S$ represents the data size in bits, and $C$ denotes the required central processing unit's (CPU) cycles for task execution in the CPU. We define a binary variable $\alpha_{n,m}(t)$, which is one if user $n$ offloads to UAV $m$ at time slot $t$, and equals zero otherwise. It is assumed that each user can offload at most one task per time slot, while each UAV $m$ is allowed to process a maximum of one task per time slot. Subsequently, we have the following constraints
\begin{align}
    \sum_{m=1}^{M} \alpha_{n, m}(t) \leq 1, \quad \forall n,t, \\
    \sum_{n=1}^{N} \alpha_{n, m}(t) \leq 1, \quad \forall m,t. 
\end{align}
Offloading the task of user $n$ to UAV $m$ at time slot $t$ is only possible if $\delta_{n, m}(t)=1$. Subsequently, we should have
\begin{equation}
    \alpha_{n, m}(t) \leq \delta_{n, m}(t).
\end{equation}
To prevent collisions between UAVs, a minimum separation distance $d_{min}$ is imposed between any two UAVs, ensuring safe operation:
\begin{equation}
    \left\|\mathbf{q}_m(t)-\mathbf{q}_{m^{\prime}}(t)\right\|_2 \geq d_{\min }, \quad \forall m \neq m^{\prime}, t.
\end{equation}

Given that the energy expenditure of the UAVs is pivotal in guaranteeing a satisfactory long service time before the UAV comes to the base for recharging, we mathematically define this quantity next. It is assumed that users communicate with their respective UAVs through orthogonal frequency-division multiple access (OFDMA) and on separate subcarriers, ensuring no interference among users. Due to the high altitude of the UAVs, LoS path is dominant, thus minimizing the impact of other factors such as small-scale fading and shadowing. Subsequently, they are not considered. This is a very common assumption in UAV trajectory design problems. As a result, the communication link between user $n$ and UAV $m$ follows a free-space path loss model. Therefore, the channel power gain between UAV $m$ and user $n$ is computed as
\begin{equation}
    h_{n, m}(t) = \frac{G_0}{\left[d_{n, m}(t)\right]^2},
\end{equation}
where $G_0$ represents the channel power gain at the reference distance of $1$ meters. If user $n$ transmits with power $P_n$ over a bandwidth $B$, and the additive white Gaussian noise (AWGN) power at UAV $m$ is given by $\sigma_m^2$, then the capacity of the link between user $n$ and UAV $m$ is given by
\begin{equation}
    r_{n,m}(t) = B \log_2\left(1 + \frac{P_n h_{n,m}(t)}{\sigma_m^2}\right).
\end{equation}

The backhaul links that connect neighboring UAVs is assumed to be perfect with negligible power consumption. This is reasonable, as neighboring UAVs have the same altitude $H$, and are in general close to one another and incur the free space path loss exponent. Hence, they can communicate with low power. For the first unprocessed task $u$ of user $n$ with data size $S_{n,u}$ bits, the transmission time $\tau_{n,u,m}(t)$ for offloading to UAV $m$ becomes
\begin{equation}
    \tau_{n,u,m}(t) = \frac{S_{n,u}}{r_{n,m}(t)}.
\end{equation}
Then, the energy consumption of the UAV for receiving this task's data becomes
\begin{equation}
    E_{n, u, m}^{r}(t) = P_m^r \times \tau_{n,u,m}(t) = P_m^r \times \frac{S_{n, u}}{r_{n, m}(t)} ,
\end{equation}
where $P_m^r$ denotes the power consumed by UAV $m$ circuits to receive data from a user. The total energy consumption of UAV $m$ equals the sum of four components: (i) hovering in the air, (ii) flying to the designated location, (iii) receiving data from users, and (iv) processing users' data. The hovering energy consumption for UAV $m$ over a single time slot with duration $\Delta t$ is represented by $E_m^{h}(t)$ and is computed as \cite{chen2022deep}
\begin{equation}
    E_m^{h}(t) = P^{h} \times \Delta t,
\end{equation}
where $P^h$ denotes the needed power for hovering. The consumed energy for flying depends on UAV speed or $\|\mathbf{v}_m(t)\|$ and is given as \cite{chen2022deep}
\begin{equation}
    E_m^{f}(t):=P^{f}\left(\left\|\mathbf{v}_m(t)\right\|\right) \times \Delta t=P^{f}\left(\left\|\mathbf{q}_m(t+1)-\mathbf{q}_m(t)\right\|_2\right)
\end{equation}
To process task $u$ of user $n$ with $C_{n,u}$ CPU cycles, it is assumed that the UAV $m$ consumes an energy per CPU cycle given by $\kappa_m$. Subsequently, the processing energy becomes
\begin{equation}
    E_{n, u, m}^{p}(t) = \kappa_m \times C_{n,u}.
\end{equation}
The total energy consumption of the UAV swarm at time slot $t$ is given by
\begin{align}
    E_{\rm total}(t) &= \sum_{n=1}^{N} \sum_{u=1}^{U_n} \sum_{m=1}^{M} \alpha_{n, m}(t) \left[ E_{n,u,m}^{r}(t) + E_{n,u,m}^{p}(t) \right]
    \notag \\ 
    &\quad + \sum_{m=1}^{M} \left[ E_{m}^{f}(t) + E_{m}^{h}(t) \right].
\end{align}
The total number of processed tasks in the system at time slot $t$ is given by
\begin{equation}
    L_{\rm pt}(t) = \sum_{n=1}^{N} \sum_{m=1}^{M} \alpha_{n, m}(t).
\end{equation}
The objective function we try to minimize is a weighted sum of total energy consumption of the UAV swarm in an episode and the negative of the total number of processed tasks during the episode. For a given time slot $t$, we define
\begin{equation}
    \Psi(t) := w_1 E_{\rm total}(t) - w_2 L_{\rm pt}(t).
\end{equation}
The non-negative weights $w_1$ and $w_2$ can be adjusted based on the specific setup and MEC provider priorities. Modifying these weights allows for a greater emphasis on either energy consumption or number of processed tasks. We minimize this objective over UAVs trajectories and the users to UAVs assignments. The formulated optimization problem becomes
\begin{align}
    \underset{\mathbf{q}, \boldsymbol{\alpha}}{\min} \quad & \sum_{t=1}^{T} \Psi (t) \label{eq:opt} \\
    \text{subject to:} \quad
    & \left\|\mathbf{q}_m(t+1)-\mathbf{q}_m(t)\right\|_2 \leq V_{\max } \Delta t, \quad \forall m,t, \tag{\ref{eq:opt}a} \label{eq:opt_a} \\
    & \alpha_{n, m}(t) \leq \delta_{n, m}(t), \quad \forall n, m, t, \tag{\ref{eq:opt}b} \label{eq:opt_c} \\
    & \sum_{m=1}^{M} \alpha_{n,m}(t) \leq 1, \quad \forall n, t, \tag{\ref{eq:opt}c} \label{eq:opt_d} \\
    & \sum_{n=1}^{N} \alpha_{n,m}(t) \leq 1, \quad \forall m, t, \tag{\ref{eq:opt}d} \label{eq:opt_e} \\
    & \left\|\mathbf{q}_m(t)-\mathbf{q}_{m^{\prime}}(t)\right\|_2 \geq d_{\min }, \quad \forall m \neq m^{\prime}, \forall t, \tag{\ref{eq:opt}e} \label{eq:opt_f} \\
    & 0 \leq x_m(t), y_m(t) \leq L, \quad \forall m, t, \tag{\ref{eq:opt}f} \label{eq:opt_g} \\
    & \alpha_{n,m}(t)\in\{0,1\}, \quad \forall n, m, t, \tag{\ref{eq:opt}g} \label{eq:opt_h} 
\end{align}
where $\mathbf{q}$ is the shorthand notation for UAV positions $\{\mathbf{q}_m(t)\}$ for all $m\in\mathcal{M}, t\in\mathcal{T}$, and $\boldsymbol{\alpha}$ is the shorthand notation for UAV to user associations $\{\alpha_{n,m}(t)\}$ for all $n\in\mathcal{N}, m\in\mathcal{M}, t\in\mathcal{T}$. As for the constraints, (\ref{eq:opt_a}) enforces the UAV maximum physically achievable movement per time slot, (\ref{eq:opt_c}) ensures that a UAV only process tasks from users in its coverage area, (\ref{eq:opt_d}), and (\ref{eq:opt_e}) enforce the task offloading constraints, (\ref{eq:opt_f}) represents the UAVs collision avoidance constraint, (\ref{eq:opt_g}) ensures that UAVs do not leave the designated square area, and \eqref{eq:opt_h} enforces the fact that partial offloading is not allowed. It means that either the complete task should be offloaded or no offloading occurs.

The optimization problem (\ref{eq:opt}) is very challenging to solve due to several reasons. As the first issue, this problem is a mixed integer nonlinear program (MINLP) with both an integer constraint in \eqref{eq:opt_h} and a continuous but non-convex constraint in (\ref{eq:opt_f}). Certain techniques such as successive convex approximation (SCA) can be utilized to replace the original constraints with continuous and convex surrogates. These methods can guarantee convergence to sub-optimal solutions under certain assumptions on the surrogates. However, achieved sub-optimal points may have poor qualities compared to the global optimum. While DRL methods do not have convergence guarantees from a rigorous mathematical perspective, numerical evaluations have revealed that they usually converge and can reach sub-optimal points with satisfactory performance. 

Proceeding with the second issue, the major challenge is not convexity. Indeed, certain parameters such as users' future locations, and subsequently their future channel power gains are not known in advance. Users move at every time slot and even though they may have been located and served before, relocating them for offloading future tasks can be difficult. Indeed, we are facing an optimization problem with partial information and some missing parameters. This means that we should solve this optimization problem in an online fashion and by predicting some unavailable parameters. 

Third challenge pertains to the distributed nature of available information. Every UAV can only acquire local information within its sensing range, and does not have a clear global picture of the environment. However, to process available tasks in an energy efficient 
manner, other UAVs observations and decisions are required. One option is to transmit all 
UAVs observations to a central unit that jointly optimizes a global objective over all UAVs decisions. However, existence of a central entity leads to many limitations that were alluded to in the Introduction. Subsequently, we advocate a fully decentralized but cooperative setup, which 
mitigates all the centralized scheme deficiencies. Under our proposed approach, every UAV makes its own 
decisions in an online fashion with its limited amount of knowledge. However, it benefits from other UAVs 
observations by communicating with its immediate neighbor UAVs before making its decision. We deploy several GAT layers to ensure efficient and satisfactory diffusion of information from UAVs several hops away in addition to nearby UAVs. Subsequently, a more accurate global view of the environment becomes available to each UAV, which will help it make more accurate decisions later on. Numerical results will reveal that we outperform existing semi-centralized approaches such as MADDPG. We will introduce our algorithm next.

\section{Proposed GAT-based EPS-PPO}
\label{sec:algorithm}

Firstly, we define a partially observable Markov decision process (POMDP) that models the optimization problem \eqref{eq:opt}. Then, we introduce our proposed fully decentralized DRL algorithm in a glance. Finally, we elaborate on each block of the algorithm and provide the full details. 

\subsection{Fitting a POMDP to the Optimization Problem}
Here, we develop the partially observable Markov decision process (POMDP) framework that effectively captures the sequential nature of UAV trajectory optimization and its association with users, where each time slot is represented as a state transition. We assume that the UAV $m$ obtains an observation vector ${\mathbf o}_m(t)$ at time slot $t$. Then, it communicates with neighboring UAVs to attain a better global picture. Given its own observation and observations from its neighbors, UAV $m$ selects an action vector $\mathbf{a}_{m}(t)$, and subsequently receives a scalar reward $r_{m}(t)$. Following this process, and given the decisions of all the UAVs, the environment evolves to a new state. The task is episodic, meaning that after $T$ time slots the POMDP ends. Still, we can select $T$ arbitrarily large to model an infinite horizon problem. Below, the core components of the POMDP are defined.
\begin{enumerate}
    \item{\textbf{Global State:}} The global state includes the position of all users in addition to the number of their remaining tasks and the tasks' size/complexity specifications. In addition, the global state contains the current position and battery level of all the UAVs. Given that this global information is not available to any individual UAV, we are facing a POMDP. Hence, each UAV's local observation is defined next.
    \item{\textbf{Local Observation:}} The local observation $\mathbf{o}_m(t)$ of UAV $m$ of the global state includes the UAV's location $[x_m(t), y_m(t), H]^T$, the remaining battery level $b_m(t)$, its coverage radius $R_{\rm cov}$, and its communication radius $R_{\rm com}$. If UAV $m$ has UAV $j$ as an immediate neighbor, the observation of UAV $m$ is expanded to include the battery status of UAV $j$, or $b_j(t)$, and the distance between UAV $m$ and UAV $j$, given by $d_{m, j}(t)$. This information helps prevent collisions and provides insights into the operational status of the neighboring UAV based on its battery level. For each user $n$ within the coverage area $R_{\rm cov}$ of UAV $m$, the observation of UAV $m$ also includes the user's location $\mathbf{q}_n(t) = [x_n(t), y_n(t)]^T$, and the number of remaining tasks assigned to that user. This allows the UAV to recognize the users it can currently serve, and efficiently manage task allocation. Finally, the observation includes a grid map, where visited grid points are labeled as one and unvisited grid points are labeled as zero, and the UAV's position on this map is also marked. This map enables the UAV to identify unexplored regions and optimize its trajectory to improve coverage and exploration efficiency.
    \item{\textbf{Possible Actions:}} At each time slot $t$, UAV $m$ performs action $\mathbf{a}_m(t)$ from its action space. The available actions are the velocity of UAV and the user to be served next.The movement action includes two continuous variables $\Delta x_m(t)$ and $\Delta y_m(t)$, which specify the UAV's movement along the $x$-axis and $y$-axis, respectively. The action related to the user being served is discrete, allowing the UAV to select a user from its list of the users within its coverage area. If $\alpha_{m,n}(t) = 1$ for a given user $n$, then user $n$ will be served next.
    \item{\textbf{Reward:}} To model the optimization problem in \eqref{eq:opt}, we bring up the constraints \eqref{eq:opt_f} and \eqref{eq:opt_g} into the objective and penalize them by the value $\lambda_m$. Hence, the reward for UAV $m$ at time slot $t$ is given by
\begin{align} \label{reward}
    r_m(t) &= -  \Psi (t) - \lambda_m\Big(1-\mathbbm{1}_{\left(0\leq x_m(t),y_m(t)\leq L\right)}\nonumber \\ & ~~\times \mathbbm{1}_{\left( \left\|\mathbf{q}_m(t)-\mathbf{q}_{m^{\prime}}(t)\right\|_2 \geq d_{\min },~\forall m\neq m'\right)}\Big).
\end{align}
Here, $\mathbbm{1}$ denotes the indicator function which is one if its logical argument is true and zero otherwise. The scalar $\lambda_m>0$ is the penalty administered to UAV $m$ if it flies out of the boundaries or if it collides with another UAV. It should be noted that provided $\lambda_m$ is selected large enough, the constrained optimization problem is equivalent to the penalized optimization problem. It is worth mentioning that when $\lambda_m$ is the same for all UAVs, and all UAVs are penalized by the same amount when any two arbitrary UAVs collide, we obtain the fully cooperative setting. In \eqref{reward}, the first term is the shared reward for all the UAVs, while the second term pertains to UAV $m$
only. This means that we may need to consider a Markov decision game (MDG) instead of an MDP. While the individual UAV rewards are not equal, with enough learning episodes, the UAVs realize how to avoid the penalty. It needs to be mentioned that UAVs' objectives are aligned in the sense that every collision will incur a penalty for the two UAVs involved, while other UAVs will not receive a penalty. The two UAVs involved in the collision learn to move away from each other and avoid future penalties, thus increasing their reward. With zero penalty, the UAV rewards will be exactly equal. Subsequently, the defined POMDP can be solved by the fully cooperative multi-agent setup. 
\end{enumerate}

Our goal is to solve this POMDP via MADRL approach. Available multi-agent approaches, differ vastly in details. One simple approach is to let every agent run an independent DRL algorithm, without any coordination with other agents. While simple and incurring zero communication overhead, this approach leads to a non-stationary environment for each agent with a high chance of divergence or oscillations. Even if this approach converges, it reaches low-quality sub-optimal points with high probability. As an example of such algorithms, independent PPO (IPPO) is analyzed by \cite{H2}. At the other extreme, one can envision a centralized server which runs a centralized DRL on behalf of all agents and then communicates the centrally determined decisions to individual UAVs. While performing satisfactorily, this approach suffers from the limitations that a central entity brings, which have been mentioned before. In between these two extremes, the CTDE approaches exist. They utilize a centralized critic with access to the full environmental state, while the actors are placed locally at individual agents. Please see MAPPO in \cite{H2} or MADDPG in \cite{lowe2017multi} for examples of this approach. Given the centralized critic, learning is centralized, but once the learning is complete, central server is disconnected and each UAV decides by itself via the local actor network. This approach can not be applied to non-stationary environments where learning is required at all time slots. Even if the learning can be stopped at some point, the centralized learning phase incurs the same limitations of a fully centralized setup. Subsequently, we advocate a fully decentralized approach with the communication occurring between neighboring UAVs only. To achieve our goal, we use some concepts in \cite{pu2022attention} but modify them to some degree. By doing so, we simultaneously reduce algorithm complexity, while maintaining the same satisfactory performance as \cite{pu2022attention}.     

\begin{figure*}[t]
    \centering
    \includegraphics[width=1\textwidth]{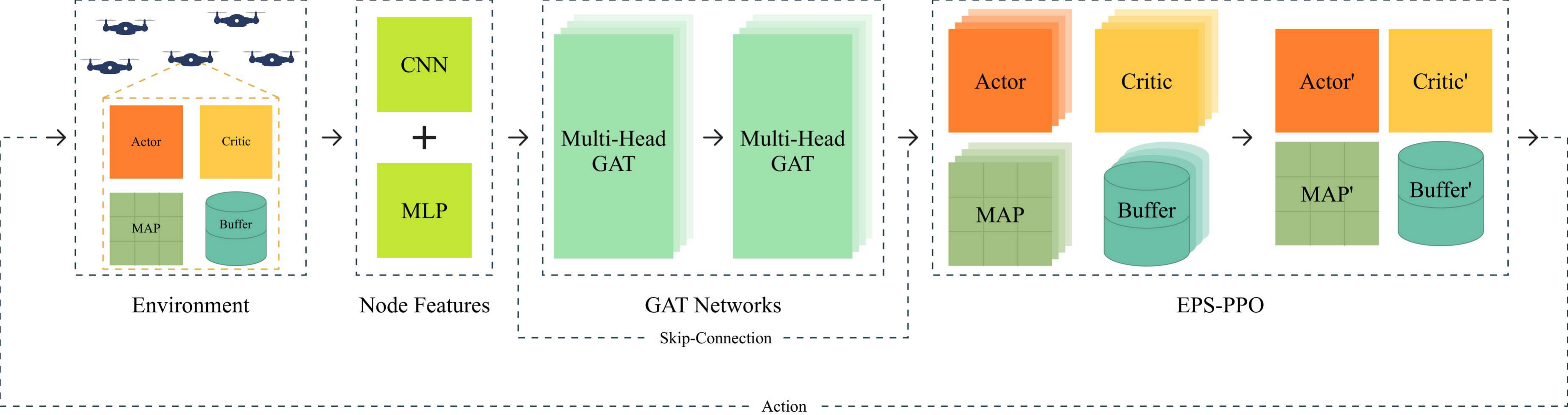} 
    \caption{Architecture of the proposed method.}
    \label{fig:algo}
\end{figure*}

\subsection{Algorithm Overview}
Fig. \ref{fig:algo} provides an overview picture of basic blocks involved in the design. At the leftmost block, every UAV obtains its individual observation which includes its map of the environment, its own status which entails position and battery level, neighboring UAVs' status, and visible users and the number of their remaining tasks. The grid map is updated through inter-UAV communication, where neighboring UAVs share their maps and mark the grid points visited by their neighbors as if visited by themselves. This prevents UAVs from redundantly exploring regions that have already been covered by other UAVs. This updated map is processed by a convolutional neural network (CNN). Parallel to CNN for processing the map, we feed other UAV’s observations such as its own status, neighboring UAVs' status, and visible users' status to a multi-layer perceptron (MLP). Afterwards, CNN and MLP output are concatenated to form the internal feature vector of the UAV. The resulting vector is fed to the GAT layers, where each UAV is represented as a node with its own feature vector. A multi-head GAT aggregates the embeddings of neighboring UAVs, combines them in a fashion that is later provided in detail, and obtains a new feature vector enriched with neighboring UAVs’ observations and information. This enhanced representation is then aggregated with the UAV's initial feature vector, which was the input to GAT. This aggregated feature vector is treated as the new observation of the UAV and fed into the local EPS-PPO algorithm, which is run locally on each UAV. In EPS-PPO, the replay buffers of the neighboring UAVs are also shared to improve the richness of the experiences available to each UAV. Given the homogeneity of the UAVs and their action and observation space, the experience sharing process helps all UAVs. At the output of the EPS-PPO algorithm, the actor network's recommended actions are taken by the UAV. 

\subsection{GAT and EPS-PPO Details}
The two blocks that bear most of the burden in our proposed approach are the GAT and EPS-PPO. We will elaborate on both below.

The GAT block consists of two GAT layers, which means that the output of the first layer is the input to the second layer. While we consider multi-head attention in each layer, we modify the well-known attention mechanism to simplify the process and prevent overfitting. Let us focus on UAV $i$ and represent itself and its immediate neighbors by the set $\mathcal{N}_i$. Furthermore, suppose that the input of UAV $i$ to the first GAT layer is denoted by $\mathbf{g}_i$. This vector will be multiplied by three different matrices, which are $\mathbf{W}_i^K,\mathbf{W}_i^Q,\mathbf{W}_i^V$ to generate key, query, and value vectors. As our modification to the general GAT, we assume that $\mathbf{W}_i^K=\mathbf{W}_i^V:=\mathbf{W}_i$, and we designate a fixed learnable vector $\mathbf{a}_i$ to represent the query vector. First, we define the combining weights $\zeta_{ij}$ as
\[
\zeta_{ij}:=\sigma\left(\mathbf{a}_i^T\mathbf{W}_i\mathbf{g}_j\right), ~~\forall j\in\mathcal{N}_i,
\]
where the function $\sigma(.)$ is a neural network activation function. It can be arbitrary, but we select it to be a leaky rectified linear unit (leaky-RELU) in our numerical analysis. Both $\mathbf{W}_i$ and $\mathbf{a}_i$ are learnable weights. The combination weight for UAV $j$ in the GAT layer for UAV $i$ is given by
\[
\alpha_{ij}=\frac{e^{\zeta_{ij}}}{\sum_{\ell\in\mathcal{N}_i} e^{\zeta_{i\ell}}}.
\]
Finally, the output of the first GAT layer for UAV $i$ is given by
\[
\tilde{\mathbf{g}}_i=\tilde{\sigma}\left(\sum_{j\in\mathcal{N}_i} \alpha_{ij}\mathbf{W}_i\mathbf{g}_j\right),
\]
where $\tilde{\sigma}(.)$ is another neural network activation function and can be same as or different from $\sigma(.)$. The same routine is repeated for the second GAT layer with $\tilde{\mathbf{g}}_i$ as the input vector of UAV $i$. To stabilize the learning process and capture multiple aspects of node relationships, multi-head attention can be employed in GAT. We utilize this multi-head attention mechanism at each GAT layers with $K$ attention heads. Subsequently, the head $k$ for the first layer of UAV $i$ GAT, evaluates 
\[
\zeta_{ij}^k:=\sigma\left(\left(\mathbf{a}_i^{k}\right)^T\mathbf{W}_i^k\mathbf{g}_j\right), ~~\forall j\in\mathcal{N}_i.
\]
Then, the combining weights for the $k$'th head of GAT layer 1 is given by
\[
\alpha_{ij}^k=\frac{e^{\zeta_{ij}^k}}{\sum_{\ell\in\mathcal{N}_i} e^{\zeta_{i\ell}^k}}.
\]
Finally, the output of the GAT layer 1 for UAV $i$ is given by
\[
\tilde{\mathbf{h}}_i=\tilde{\sigma}\left(\frac{1}{K}\sum_{k=1}^K \sum_{j\in\mathcal{N}_i} \alpha_{ij}^k\mathbf{W}_i^k\mathbf{g}_j\right),
\]
which will be the input of UAV $i$ to the second GAT layer. It should be noted that the output of the second GAT layer, i.e., $\tilde{\bar{\mathbf{g}}}_i$, for UAV $i$ is concatenated with the input to the first GAT layer, or $\mathbf{g}_i$, to emphasize the observation of UAV $i$ itself compared to further information obtained from combining neighboring UAVs information. The output of the second GAT layer will become the UAV $i$ final observation which is later fed to the local EPS-PPO of UAV $i$. It should be noted that further GAT layers can be used. They will increase complexity but can obtain information from UAVs that are several hops away in the graph. Our numerical experiments revealed that two GAT layers offer the best such trade-off.

Due to its simplicity and stability, the PPO algorithm \cite{schulman2017proximal} has become a widely adopted option in versatile applications, when a centralize approach is pursued. In the scenarios with multiple agents, semi or fully decentralized algorithms are preferred due to reasons mentioned before. Fully decentralized IPPO is simple and does not incur communication overhead but does not perform very well. MAPPO is semi decentralized and performs satisfactorily, but its learning phase is centralized, which is not very convenient due to the limitations of the centralized approach. Here, we advocate the local, and fully decentralized EPS-PPO, where each individual UAV runs on its own platform. However, both old experiences that are located in the replay buffer, and updated actor and critic weights are shared and averaged between neighboring UAVs. It needs to be noted that information sharing between neighboring UAVs occur in three places. Firstly, UAVs maps are shared before entering the GAT block. Secondly, the UAVs local information is diffused in the communication graph via the GAT layers. Third, EPS-PPO shares experiences before PPO is applied and share latest actor-critic networks' weights after every PPO update. It needs to be mentioned that the original PS-PPO does not share experiences between multiple agents. It only shares actor-critic networks' weights. While allowing to share replay buffer experiences in our EPS-PPO, the state in experiences in the replay buffer are given by the GAT processed $\mathbf{z}_m^t$, rather than the raw local observations. The GAT processed $\mathbf{z}_m^t$ contains information about other UAVs and their measurements, which were diffused through the UAV network via GAT layers. Hence, some global information is also captured in $\mathbf{z}_m^t$, which will significantly improve EPS-PPO performance. This observation amounts to the major difference between existing PS-PPO, and our proposed approach. 


\begin{algorithm}[t!]
\DontPrintSemicolon
\caption{Decentralized EPS-PPO with GAT}
\label{alg:decentralized-ppo}
\KwIn{Number of UAVs $M$, Maximum number of episodes $\mathcal{E}_{\max}$, Communication radius $R_{\rm com}$, and Coverage radius $R_{\rm cov}$.}
\Init{}{Actor and Critic parameters $\boldsymbol{\theta}_m, \boldsymbol{\phi}_m$, and replay buffers $\mathcal{D}_m$.} 
\For{{\rm episode} = 1 to $\mathcal{E}_{\max}$}{
  Reset environment. \\
  \For{$t = 1$ to $T$}{
  \If{t=1}{Each UAV collects the initial observation $\mathbf{o}_m(1)$.}
    \For{{\rm each UAV} $m$ {\rm in parallel}}{
      Give $\mathbf{o}_m(t)$ to the CNN/MLP to compute node feature $\mathbf{g}_m^t$.\\
      Feed $\mathbf{g}_m^t$ to GAT and compute GAT output $\tilde{\bar{\mathbf{g}}}_m^t$.\\
      Skip-connection: $\mathbf{z}_m^t = [\mathbf{g}_m^t \,\|\, \tilde{\bar{\mathbf{g}}}_m^t]$.\\
      Sample action $\mathbf{a}_m^t \sim \boldsymbol{\pi}_{\boldsymbol{\theta}_m}(\mathbf{a} \mid \mathbf{z}_m^t)$.\\
      Execute $\mathbf{a}_m^t$, receive $r_m^t$ and obtain $\mathbf{o}_m(t+1)$\\
      \If{$t > 1$}{Store $(\mathbf{z}^m_{t-1}, \mathbf{a}_m^t, r_m^t, \mathbf{z}^m_{t})$ in $\mathcal{D}_m$.}
       \If{$j\in\mathcal{N}_m$}{ $\mathcal{D}_m \gets \mathcal{D}_m \cup \mathcal{D}_j$.}
      Run PPO on a mini-batch from $\mathcal{D}_m$ to obtain new $\boldsymbol{\tilde{\theta}}_m$ and $\boldsymbol{\tilde{\phi}}_m$.\\
      Communicate $\boldsymbol{\tilde{\theta}}_m$ and $\boldsymbol{\tilde{\phi}}_m$ to the neighbors.\\ Evaluate
      \[
          \boldsymbol{\theta}_m \gets \frac{\sum_{j \in \mathcal{N}_m} \boldsymbol{\tilde{\theta}}_j}{|\mathcal{N}_m|}, ~~
          \boldsymbol{\phi}_m \gets \frac{\sum_{j \in \mathcal{N}_m} \boldsymbol{\tilde{\phi}}_j}{|\mathcal{N}_m|}.
      \]
     } end
      }  end
    } end \\
\KwOut{Trained local actor at UAV $m$ with policy $\boldsymbol{\pi}_{\boldsymbol{\theta}_m}$ for all $m\in\mathcal{M}$.}
\end{algorithm}

\subsection{The Complete Algorithm}
We have presented the complete pseudo-code of our proposed method in Algorithm \ref{alg:decentralized-ppo}. At the start of each episode, every UAV initializes its actor and critic network weights and sets up a replay buffer. After applying the CNN layer for encoding the map, and MLP to encode other observations, two multi-head GAT layers diffuse information from nearby UAVs. Finally, an EPS-PPO algorithm is run locally to optimize decision-making. This algorithm supports fully decentralized training and execution, making it a desirable approach specially with a large number of UAVs. unlike methods such as MADDPG, which typically assume a fixed number of agents and a centralized critic, there is no need for retraining when the number of UAVs changes. Indeed, UAVs can be further added or grounded without any modification or extra burden to the proposed algorithm, as long as the communication network remains connected. 

\section{Numerical Results}
\label{sec:results}
Here, we evaluate the performance of our proposed algorithm and compare it against an existing alternative. In addition, we demonstrate the algorithm's  scalability by varying the numbers of UAVs. In those experiments where the number of UAVs and/or users are fixed, we assume 10 UAVs and 50 users. These numbers simulate a highly dense environment in terms of number of UAVs and users. In the training phase, we run the algorithm on 900 episodes, where every episode have 80 time slots. Simulation parameters are provided in Table \ref{sim_param}.

\begin{figure}[t!]
    \centering
    \includegraphics[width=\columnwidth]{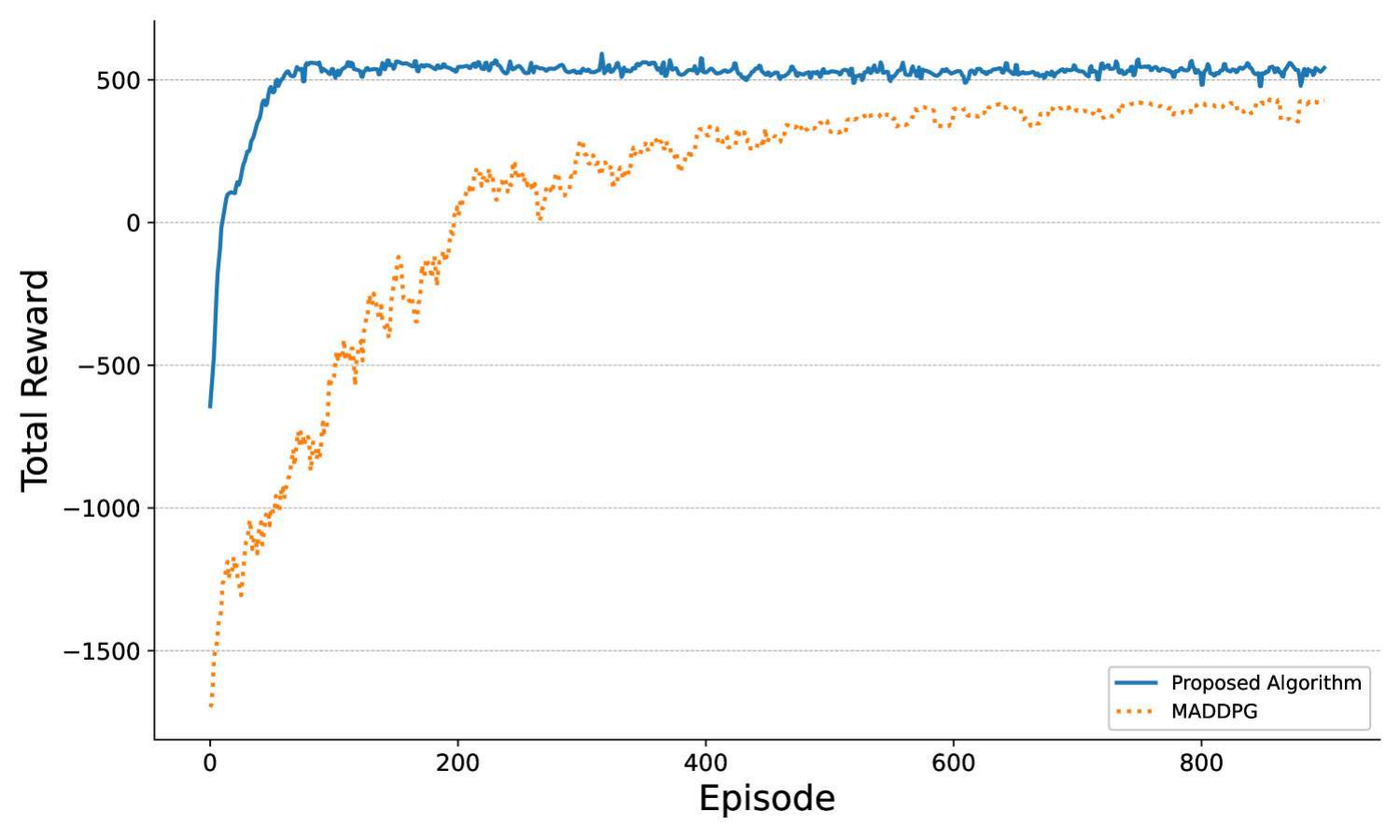}
    \caption{Sum discounted reward during training for the proposed GAT-based EPS-PPO and the MADDPG benchmark.}
    \label{fig:result_reward}
\end{figure}

Fig. \ref{fig:result_reward} shows the convergence properties of our proposed algorithm and the MADDPG benchmark. It can be observed that our proposed GAT-based EPS-PPO converges faster than MADDPG, and to a better sub-optimal point. Convergence curve of our proposed method has smaller ripples, and flattens at a higher sum discounted reward than MADDPG. This fact suggests that in addition to the operational benefits one has in a fully decentralized algorithm, performance criteria can be better as well. The weaker performance of MADDPG can be attributed to the extremely large size of the centralized critic. The inputs to the critic are the global environment state and the selected actions of all UAVs. While the global state of the environment is significantly larger than the local enriched feature vector $\mathbf{z}_m$ in size, action space of a single UAV is considerably smaller than the action space for all UAVs. Subsequently, with a large number of UAVs, it can be deduced that the size of the centralized critic can be extremely large. Thus, it will take more time to train well and avoid overfitting. This observtion explains the slower convergence rate of MADDPG. On the other hand a larger critic usually amounts to more local optima. Hence, it increases the probability of getting stuck in a poor local optimum. This fact explains the smaller discounted sum reward of the MADDPG upon convergence. One concludes that by removing the central unit, we omit a learning bottleneck. There exist significant improvements in the early episodes due to strong penalties on UAV collisions and boundary violations. As the learning process continues, these unfavorable behavior occur less often, which cause the learning curve to flatten gradually. After initial episodes, UAVs learn to avoid collisions and thus the penalty. Given zero penalty, UAVs focus on minimizing energy consumption and maximizing number of offloaded tasks, which is the observed phenomenon at later episodes.

\begin{table}[t!]
\centering
\caption{Simulation Parameters}
\label{sim_param}
\small 
\begin{tabular}{|p{4.2cm}|p{3.5cm}|} 
\hline
\textbf{Parameter} & \textbf{Value} \\
\hline
Number of UAVs ($M$) & 10 \\
\hline
Number of users ($N$) & 50 \\
\hline
Number of time slots ($T$) & 80 \\
\hline
Horizontal area size ($L$) & 250 m \\
\hline
UAV altitude ($H$) & 100 m \\
\hline
Task size ($S_{u,n}$) & [100, 200] Kb \\
\hline
Task CPU cycles ($C_{u,n}$) & [150, 200] cycles/bit \\
\hline
Minimum distance ($D_{\min}$) & 10 m \\
\hline
Coverage radius ($R_{\rm cov}$) & 25 m \\
\hline
Communication radius ($R_{\rm com}$) & 60 m \\
\hline
Bandwidth ($B$) & 10 MHz \\
\hline
Power gain ($G_0$) & -50 dB \\
\hline
Noise power ($\sigma_m^2$) & -90 dBm \\
\hline
User transmit power ($P_n$) & 0.1 W \\
\hline
UAV receiver power ($P_m^r$) & 0.1 W \\
\hline
DRL discount factor ($\gamma$) & 0.99 \\
\hline
Initial UAV battery & 100 kJ \\
\hline
CPU energy per cycle ($\kappa_m$) & $10^{-27}$ \\
\hline
Max UAV speed ($V_{\max}$) & 2 m/s \\
\hline
Max energy ($\mathcal{E}_{\max}$) & 900 \\
\hline
Hovering power ($P^h$) & 1 W \\
\hline
Flying power ($P^f$) & 10 W \\
\hline
Penalty factor ($\lambda_m$) & 500 \\
\hline
Number of neighbors ($K$) & 4 \\
\hline
Weights ($w_1 = w_2$) & 0.5 \\
\hline
\end{tabular}
\end{table}

\begin{figure}[t!]
    \centering
    \includegraphics[width=\columnwidth]{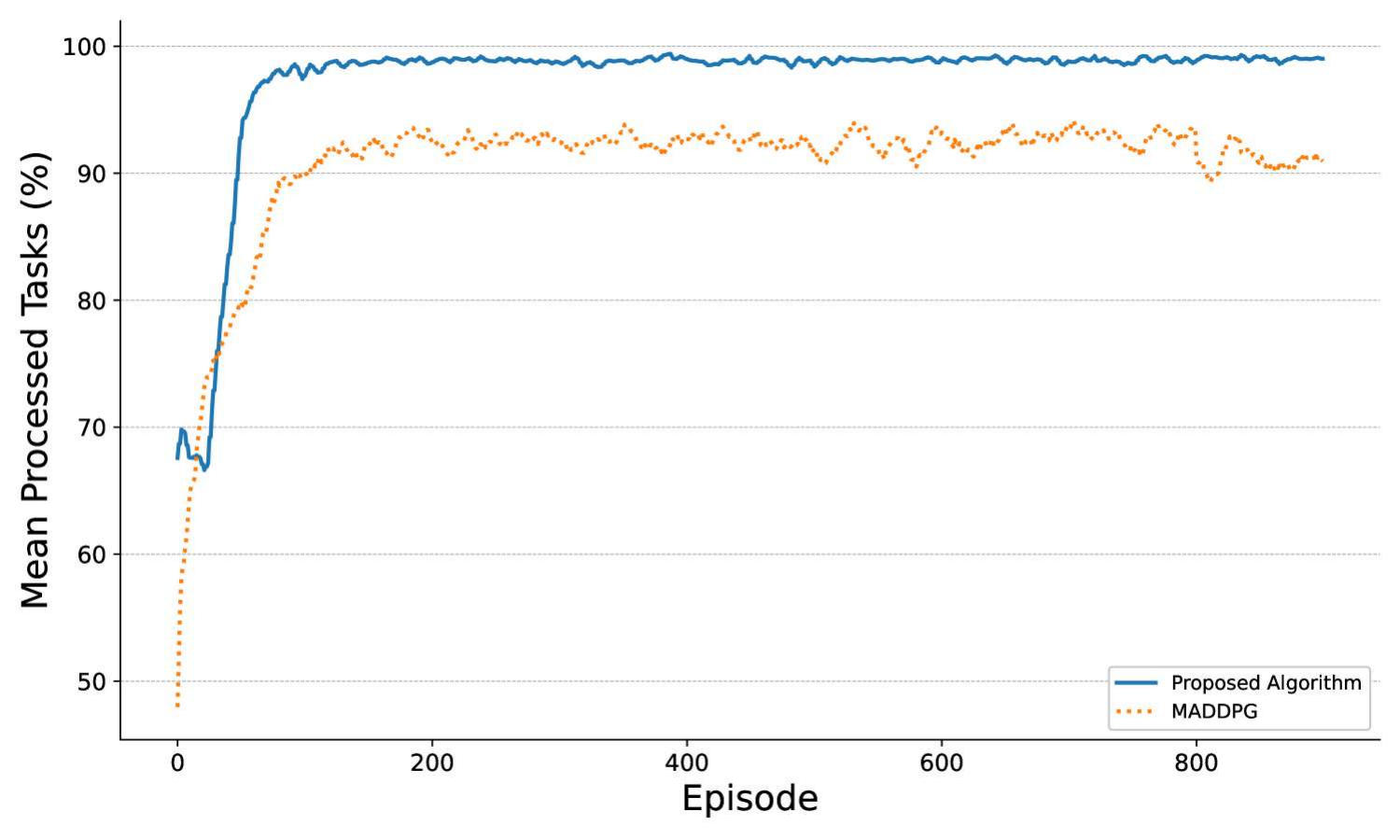}
    \caption{Task processing percentage plots for the proposed GAT-based EPS-PPO and MADDPG.}
    \label{fig:result_tasks}
\end{figure}

The proposed algorithm achieves a higher percentage of successfully processed tasks, demonstrating a more effective UAV trajectory design, and improved coordination among UAV agents. Fig. \ref{fig:result_tasks} illustrates the mean percentage of successfully processed tasks over training episodes for both our proposed GAT-based EPS-PPO and the MADDPG benchmark. Our approach shows a rapid increase in task processing percentage, stabilizing at approximately 99\% within the first 100 episodes. In contrast, MADDPG converges more slowly and stabilizes around 92–93\%, with larger fluctuations in its curve. The improved performance of our approach indicates our agents' ability to cover more areas with less energy consumption. While our proposed algorithm performs close to optimal, which amounts to 100\% successful task offloading percentage, it does not exactly reach it. This is due to user mobility and the geometric constraints of the environment. Specifically, users located in corner regions may not be serviced, since corners are risky for UAVs due to high penalty costs for boundary violations. As a result, the final mean task completion percentage is slightly smaller than 100\%.

\begin{figure}[t!]
    \centering
    \includegraphics[width=\columnwidth]{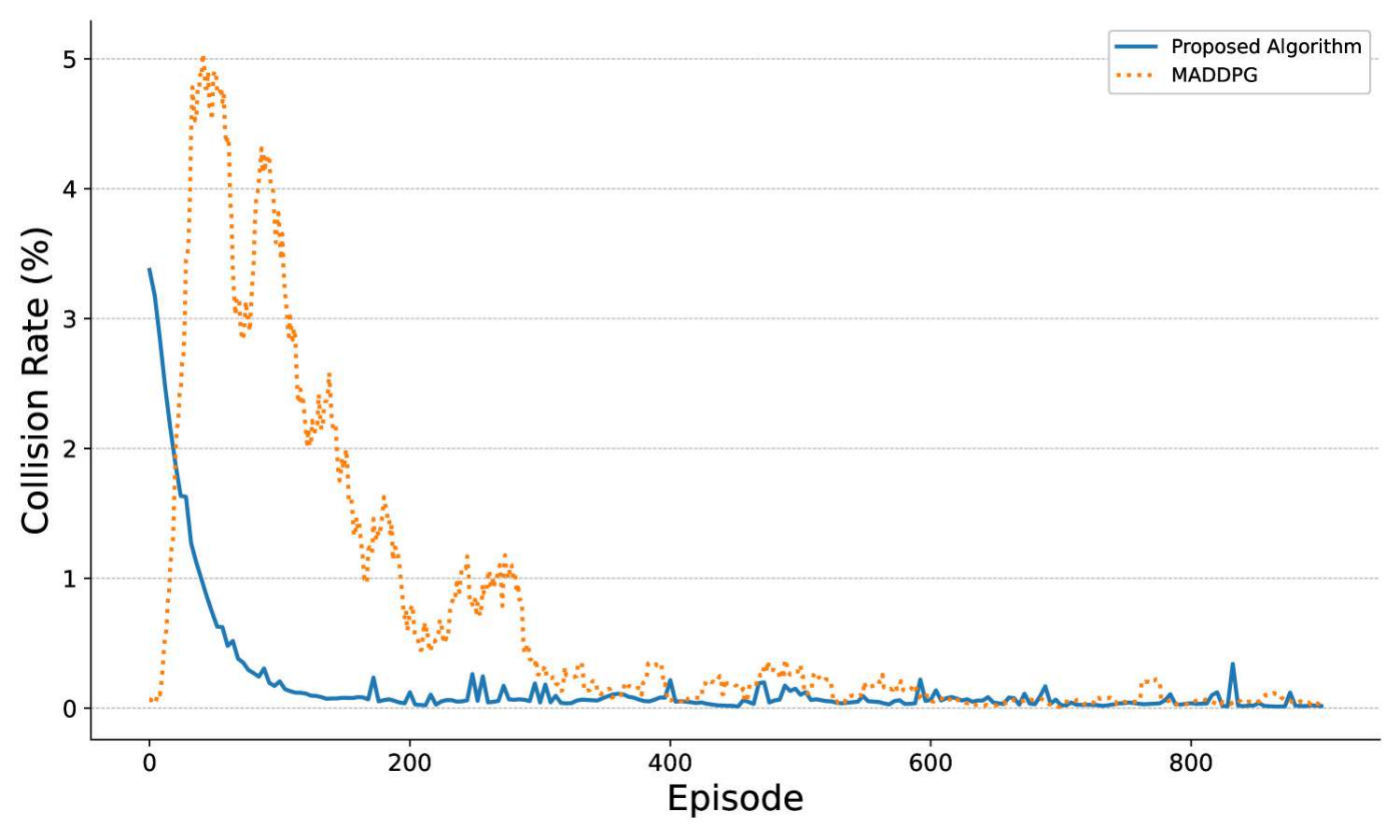}
    \caption{Collision rate over training episodes for the proposed GAT-based EPS-PPO and MADDPG.}
    \label{fig:result_collisions}
\end{figure}

Fig. \ref{fig:result_collisions} illustrates the collision rate of UAVs versus the training episodes. When training begins, the collision rate is relatively high. However, as training progresses, the collision rate drops significantly, eventually stabilizing near zero after approximately 200 episodes. This demonstrates that our algorithm effectively learns collision-avoidance strategies, resulting in safer UAV navigation within the MEC environment. It is important to note that collisions persist through the later episodes, due to the random initialization of UAV positions at the start of each episode. Occasionally, UAVs are placed close to each other, which leads to collisions in the early time slots. However, as the episode progresses, the UAVs adjust their positions, maintaining a safe distance from one another.  

\begin{figure}[t!]
    \centering
    \begin{subfigure}[b]{0.45\textwidth} 
        \includegraphics[width=\linewidth]{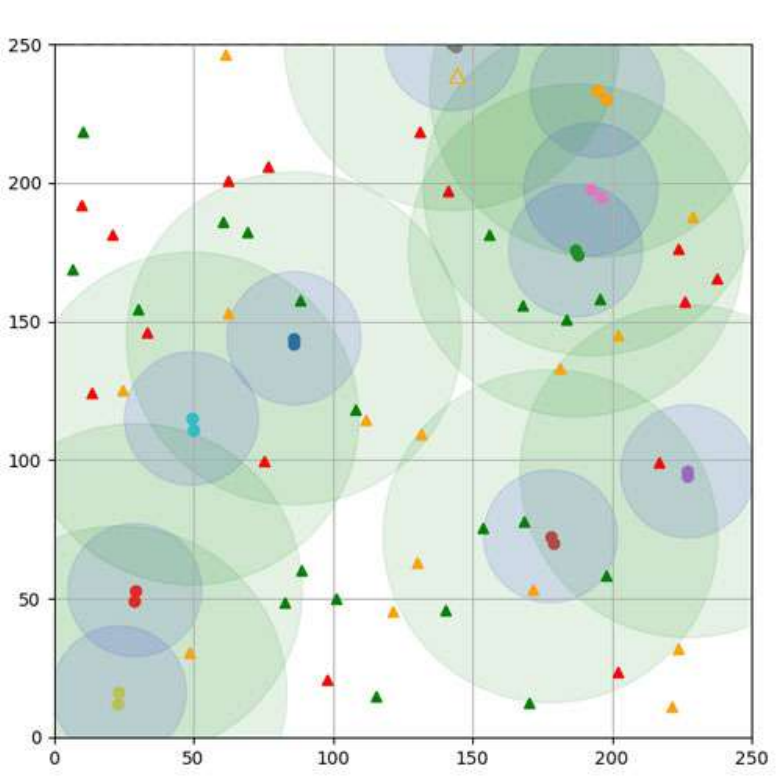} 
        \caption{}
        \label{fig:env_a}
    \end{subfigure}
    \begin{subfigure}[b]{0.45\textwidth} 
        \includegraphics[width=\linewidth]{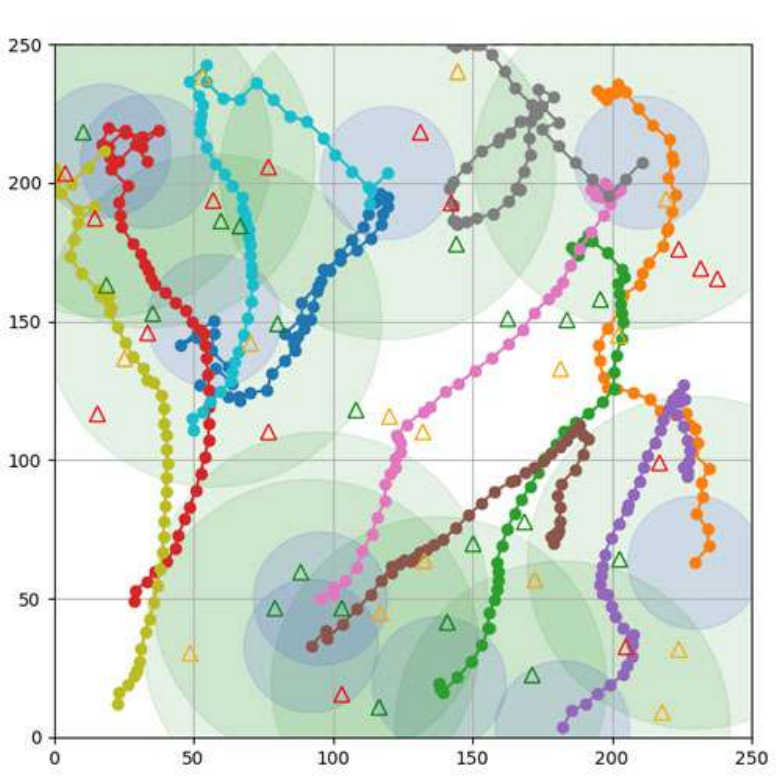} 
        \caption{}
        \label{fig:env_b}
    \end{subfigure}
     \caption{(a) Placement of the UAVs and the users at the beginning of the episode, (b) UAV trajectories at the end of the episode.}
    \label{fig:env}
\end{figure}

\begin{figure}[t!] 
        \centering  
        \includegraphics[width=0.45\textwidth]{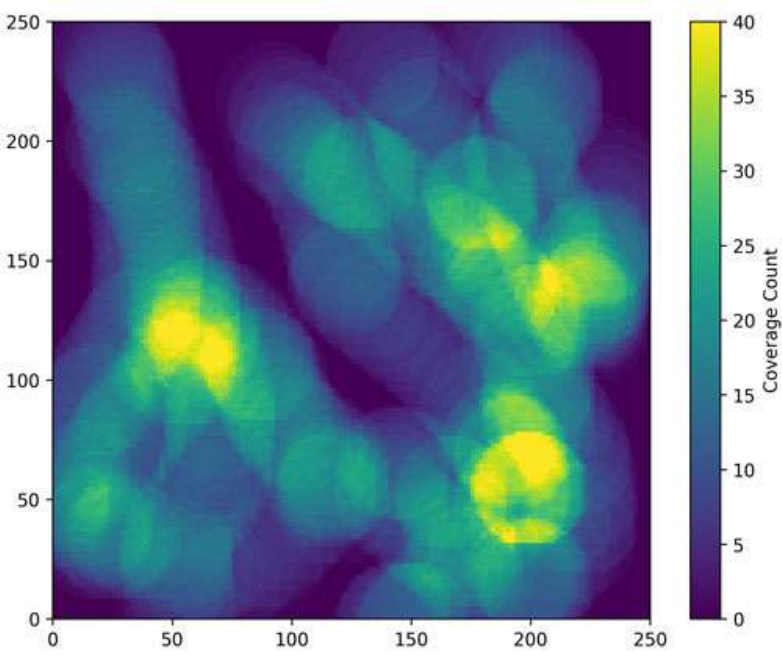} 
        \caption{Coverage heat map for Fig. \ref{fig:env_b}.}
        \label{fig:env_c}
    \end{figure}

\begin{figure}[t!]
    \centering
    \includegraphics[width=\columnwidth]{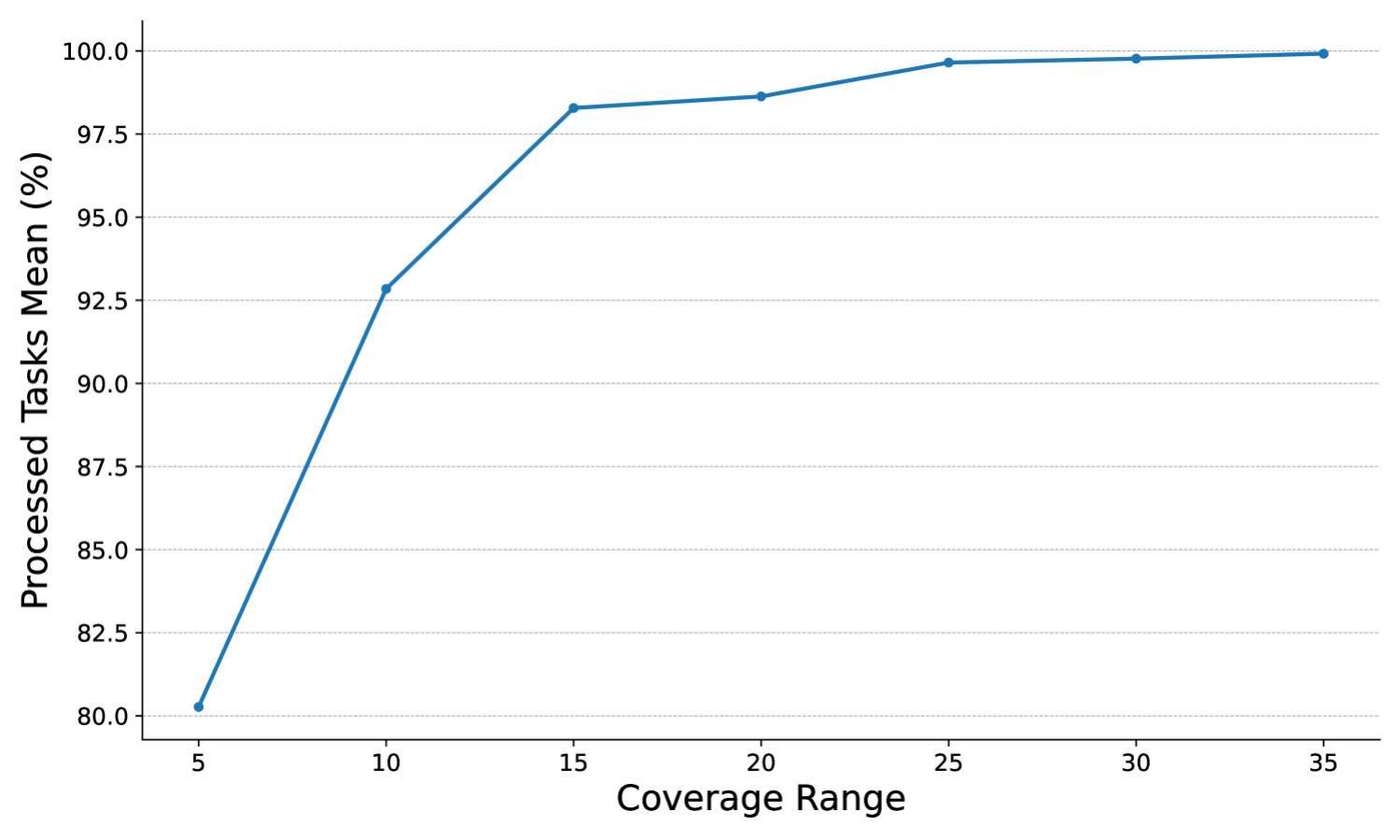}
    \caption{Effect of UAV coverage range on mean processed task percentage.}
    \label{fig:result_coverage}
\end{figure}

In Fig. \ref{fig:env}, we illustrate the performance of the trained UAVs at the beginning and the end of one episode. Users are depicted as triangles, while UAVs are represented by circles. Each UAV explores the area, while cooperating with others to service the moving users. As more UAVs are deployed in the area, their trajectories are adjusted to minimize overlap between the UAVs monitoring regions. This collaborative approach ensures comprehensive area coverage, as we see it in the heat map in Fig. \ref{fig:env_c}.

Fig. \ref{fig:result_coverage} illustrates the mean percentage of tasks processed over 400 test episodes as a function of the UAV coverage range. As the coverage range increases, the mean task processing rate improves significantly, particularly within the range of 5 to 25 meters. Beyond approximately 25 meters, there is a saturation in performance, with only marginal gains in task processing rate, approaching near 100\% efficiency. This trend indicates that while increasing the coverage range enhances task processing performance, beyond a certain point the improvements become incremental. According to the results, coverage ranges of 15 and 25 meters serve as elbow points, where performance gains begin to flatten. Based on this trade-off in Fig. \ref{fig:result_coverage}, we select a coverage range of 25 meters for our numerical comparison.

\begin{figure}[t!]
    \centering
    \includegraphics[width=\columnwidth]{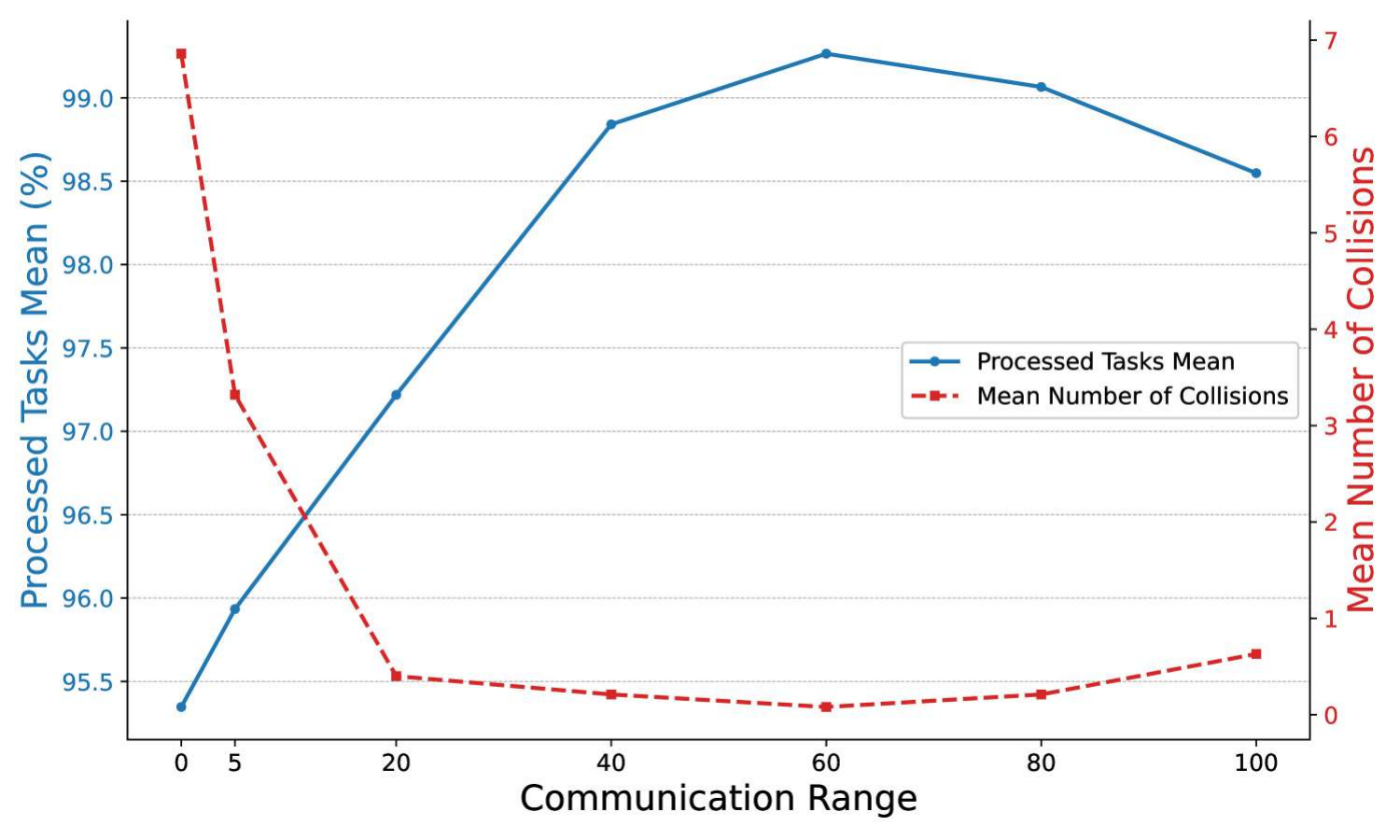}
    \caption{Effect of communication range on mean processed task percentage and average number of collisions.}
    \label{fig:result_communication}
\end{figure}

Fig. \ref{fig:result_communication} presents the effect of the communication range of UAVs on two key metrics: (i) The mean percentage of processed tasks, and (ii) the mean number of collisions. A sharp reduction in the collision rate is seen as the communication range increases from 0 to 20 meters, indicating enhanced situational awareness and inter-UAV coordination. As the communication range increases, the task processing rate gradually improves, reaching a peak of just over 99\% at approximately 60 meters. However, as the communication range extends beyond 60 meters, a slight increase in collisions, and reduction in processed tasks percentage is observed. This behavior is due to information overload, where UAVs begin to receive excessive or irrelevant data from a larger number of neighbors. The large volume of shared information can hinder decision making. Additionally, we enforce that only a limited number of neighbors should to be selected to communicate via GAT to limit hardware size and processing complexity. When a UAV has more neighbors than it can communicate with in GAT, the nearest neighbors are selected. Given this heuristic selection, some important connections may be dropped leading to reduced performance for large communication ranges.

\begin{figure}[t!]
    \centering
    \includegraphics[width=\columnwidth]{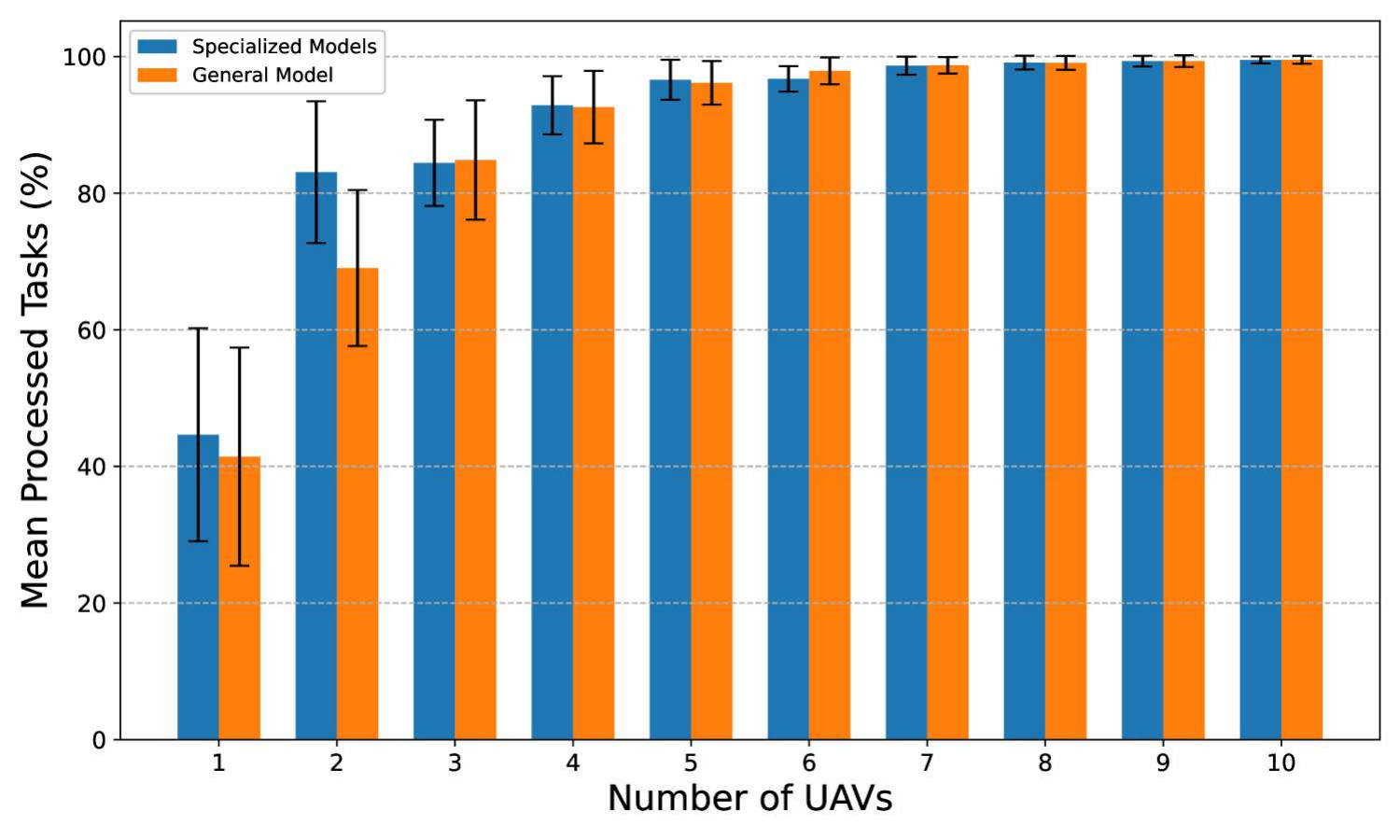}
    \caption{Comparison between general and specialized models across different numbers of UAVs.}
    \label{fig:result_different_models}
\end{figure}

Fig. \ref{fig:result_different_models} compares the performance of two similar models trained under different scenarios. The orange (second) bars represent the scenario where 7 UAVs are deployed for training, but test is carried out with a different number of UAVs. It is referred to as the general training setup. The blue (first) bars represent specialized training, where training and test are carried out over an equal number of UAVs. Clearly, it is expected that first bars be higher as train and test setups are similar in terms of number of UAVs. For a smaller UAV fleet, the specialized training significantly outperforms the general training approach. This is due to the fact that $7$ UAVs can cover the whole region with less movement hence they assign a higher priority to energy consumption. This is not the case for a small number of UAVs as they need to move around a lot to cover for their small number. Hence, the near optimal strategy differs significantly between these two setups. However, prioritizing energy consumption is a valid strategy for a large number of UAVs. Subsequently, the gap between the two bars disappears for a large number of UAVs during test. The general training setup is desirable due to its scalability, meaning that you do not have to separately train the UAVs as their number changes. Hence, it is very efficient in terms of processing complexity. This capability is particularly beneficial in dynamic environments, where the number of UAVs may vary due to fluctuating network demands or UAV availability.

\begin{figure}[t!]
    \centering
    \includegraphics[width=\columnwidth]{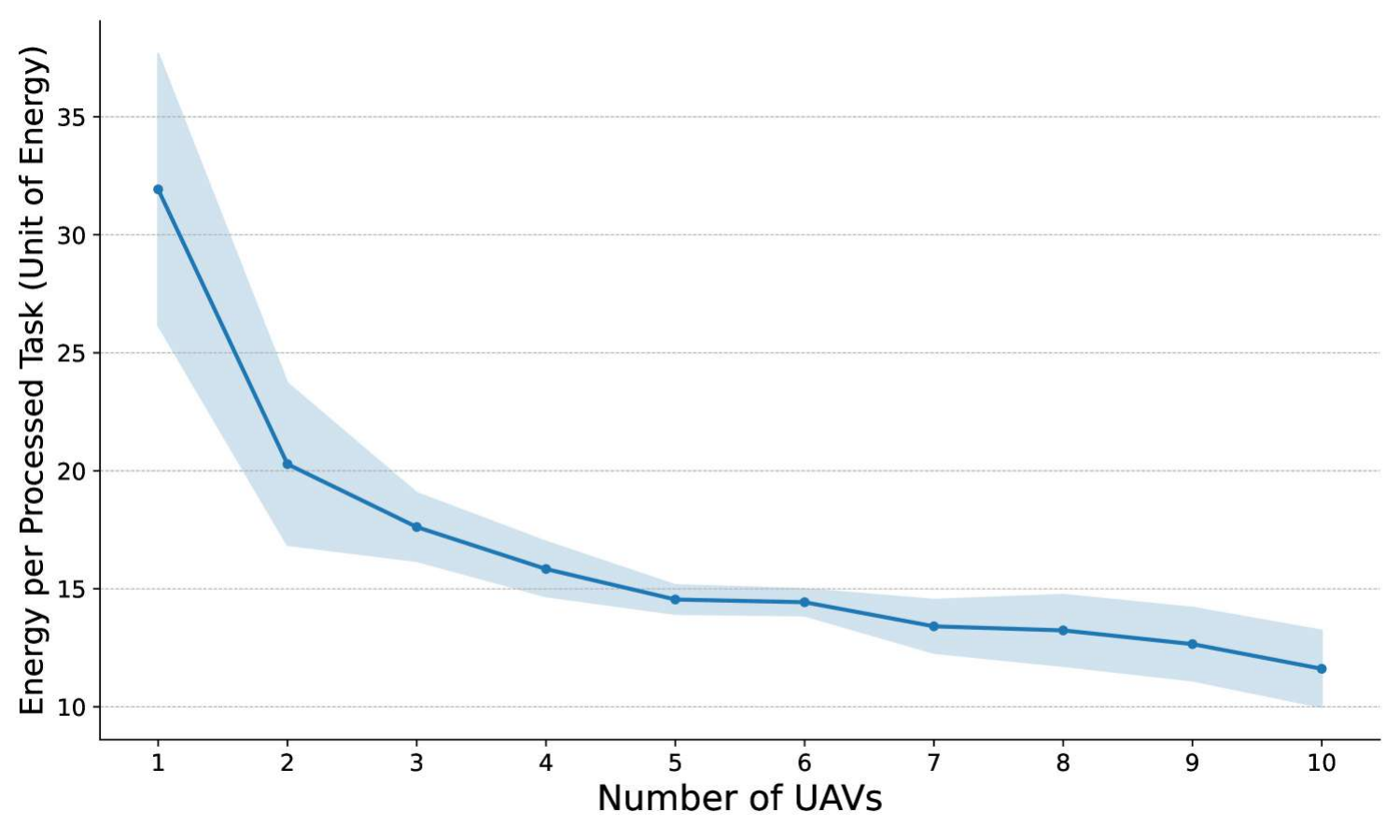}
    \caption{Energy per processed task versus number of UAVs.}
    \label{fig:result_ener_per_task}
\end{figure}

Fig. \ref{fig:result_ener_per_task} depicts the relationship between the number of UAVs and the energy consumed per processed task. As the number of UAVs increases, the average energy required to complete each task gradually decreases. So the workload is reduced per UAV, which enables more efficient trajectory planning and lower individual energy expenditure. The sharpest drop in energy consumption is observed between 1 and 3 UAVs, highlighting the significant benefit of cooperative behavior in sparse deployments. In smaller fleets, the lack of coordination forces individual UAVs to search larger areas for tasks, increasing energy consumption. In contrast, as the fleet grows, UAVs benefit from greater task visibility and shared user information using GAT, reducing the need for extensive searching and flight time. Beyond 5 UAVs, the energy consumption continues to shrink, but at a slower rate, as fleet size increases. The shaded region represents variance, which also narrows with an increasing number of UAVs. The narrowing pattern indicates more stable and consistent energy efficiency in larger UAV swarms.

\begin{figure}[t!]
    \centering
    \includegraphics[width=\columnwidth]{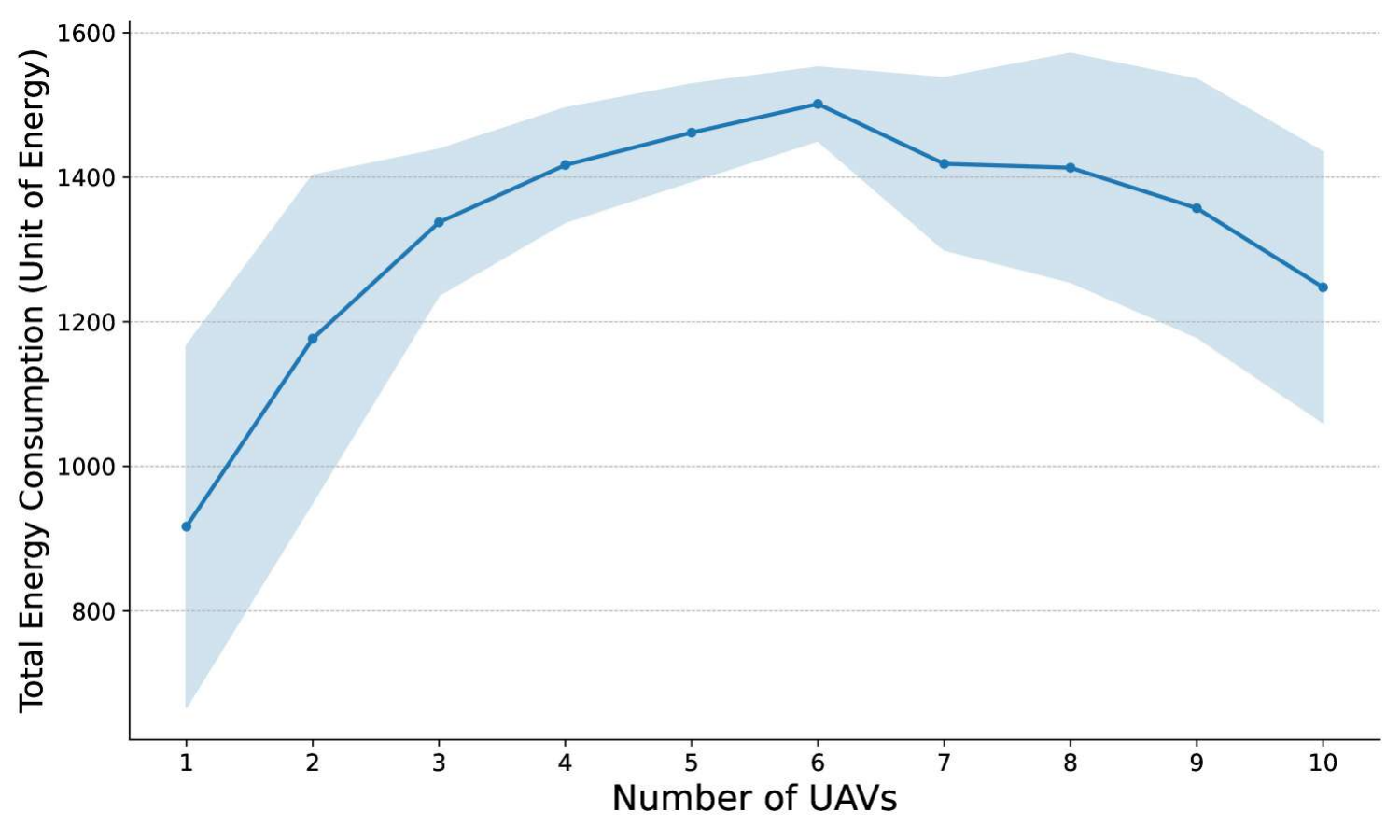}
    \caption{Total energy consumption versus number of UAVs.}
    \label{fig:result_total_energy}
\end{figure}

Fig. \ref{fig:result_total_energy} illustrates how total energy consumption varies with the number of UAVs. Initially, total energy usage increases as more UAVs are added. This is driven by higher overall activity and greater task processing capacity. However, after reaching a peak around 6 UAVs, the total energy consumption declines gradually. This trend suggests that while adding more UAVs to a small fleet results in greater collective energy use, the system becomes more energy efficient as the fleet size grows. This phenomenon occurs, because task processing workload is better distributed among UAVs and long distance flights become less frequent. Subsequently, UAVs perform much more hovering and less movement which reduces total consumed energy, in spite of more active UAVs flying simultaneously. Another notable observation is the wider variance in energy consumption for smaller UAV fleets, which indicates lower system stability and greater workload fluctuations in small fleet deployments. As the swarm size increases, this variance narrows similar to Fig. \ref{fig:result_ener_per_task}.

\section{Conclusion}
\label{sec:conclusions}

In this work, a fully decentralized multi-agent deep reinforcement learning framework was presented to optimize UAV trajectories, task offloading, and energy consumption in UAV-enabled MEC systems. Unlike current centralized and semi-centralized approaches, our method enables UAVs to learn optimal policies on their own through only local coordination and interaction, thereby eliminating the need for a central control unit. Our proposed approach led to significant improvements in energy consumption, number of processed tasks, collision rates, and convergence speed compared to the  MADDPG approach. It was concluded that the proposed GAT-based EPS-PPO outperforms DDPG, and similar methods, in performance, while simultaneously ensuring adaptability, scalability, and robustness to failures in dynamic and uncertain MEC environments.

%
%
%
%
%


%
%
%
%
%

\bibliographystyle{unsrt}
\bibliography{Myrefs} 

\end{document}